\documentclass[twocolumn,showpacs,preprintnumbers,amsmath,amssymb,prl]{revtex4}

\usepackage{graphicx}
\usepackage{dcolumn}
\usepackage{bm}
\usepackage[mathscr]{eucal}

\newcommand{\st}[2]{\stackrel{\mbox{\tiny (#1)}}{#2}\hspace{-0.1cm}}

\begin{document}

\title{Magnetic field generation from cosmological perturbations}

\author{Keitaro Takahashi$^{(1)}$, Kiyotomo Ichiki$^{(2)}$, Hiroshi Ohno$^{(3)}$
and Hidekazu Hanayama$^{(2,4)}$}
\affiliation{$^{(1)}$ Department of Physics, Princeton University,
Princeton, NJ 08544, USA}
\affiliation{$^{(2)}$
National Astronomical Observatory of Japan, Mitaka, Tokyo 181-8588, Japan}
\affiliation{$^{(3)}$ Laboratory, Corporate Research and Development Center,
Toshiba Corporation, 1 Komukai Toshiba-cho, Saiwai-ku, Kawasaki 212-8582, Japan}
\affiliation{$^{(4)}$ Department of Astronomy, University of Tokyo,
7-3-1 Hongo, Bunkyo-ku, Tokyo 113-0033, Japan}
\hspace{1cm}
\date{\today}

\begin{abstract}
In this letter, we discuss generation of magnetic field from cosmological perturbations.
We consider the evolution of three component plasma (electron, proton and photon)
evaluating the collision term between elecrons and photons up to the second order.
The collision term is shown to induce electric current, which then generate magnetic field.
There are three contributions, two of which can be evaluated from the first-order quantities,
while the other one is fluid vorticity which is purely second order. We estimate
the magnitudes of the former contributions and shows that the amplitude of the produced
magnetic field is about $\sim 10^{-19} {\rm G}$ at 10Mpc comoving scale at the decoupling.
Compared to astrophysical and inflationary mechanisms for seed-field generation,
our study suffers from much less ambiguities concerning unknown physics and/or processes.
\end{abstract}

\maketitle

\section{Introduction}

There are convincing evidences that imply existence of substantial magnetic fields
in various astronomical objects. Not only galaxies but systems with even larger scales,
such as cluster of galaxies and extra-cluster fields, have their own magnetic fields
(for a review on cosmological magnetic fields, see e.g., \cite{Widrow02}).
Conventionally the magnetic fields in galaxies, and possibly in clusters of galaxies,
are considered to have been amplified and maintained by dynamo mechanism.
However, the dynamo mechanism needs the seed magnetic field and does not explain the origin
of the magnetic fields.

There have been many attempts to generate the seed fields. One of the approaches
to this problem is to generate magnetic fields astrophysically, often involving
the Biermann mechanism \cite{Biermann50}. This mechanism has been applied to various
systems: large-scale structure formation \cite{Kulsrud97}, ionizing front \cite{Gnedin00},
protogalaxies \cite{Davies00} and supernova remnant of the first stars \cite{Hanayama05}.
These studies showed the possibilities of magnetic-field generation with amplitudes of order
$10^{-16} \sim 10^{-21} {\rm G}$, which would be enough for the required field
$\sim 10^{-20} {\rm G}$. 

On the other hand, cosmological origins, which are often concerned with inflation,
can produce magnetic field with coherence lengths of much larger scales, which is typically
the horizon scale or possibly super-horizon scale
\cite{Turner88,Bamba04,Ashoorioon04,Bertolami99,Bertolami05}. Also they can often produce
fields with a wide range of length scales, while astrophysical mechanisms can produce fields
only with their characteristic scales. For a constraints on magnetic field with
cosmological scale, see \cite{Yamazaki04}.

In this Letter, we consider magnetic-field generation from cosmological perturbations
after inflation. Around and after the decoupling, coupling among charged particles
and photons become so weak that electric current can be induced by the difference
in motions of protons and electrons. This electric current leads to generation of
magnetic fields. It is well known that vorticity of plasma produces such electric current
\cite{Harrison70,Lesch95,Hogan00,Berezhiani04,Matarrese04,Gopal04}. However, because vorticity
is not produced at the first order in cosmological perturbations, we must study
the second order. We will consider equations of motion for protons, electrons and
photons separately up to the second order, although equation of motion for photons
does not appear explicitly. To study electric current appropriately, we must treat
the three components separately \cite{Gopal04}. Furthermore, we will
evaluate the collision term between electrons and photons, which is dominant and
essential for the magnetic-field generation, up to the second order. From the equations
of motion for protons and electrons, we obtain a generalized Ohm's law, which, combined
with the Maxwell equations, leads to an evolution equation for magnetic field.
Our study is based on the cosmological perturbation theory, which is highly successful
in the anisotropies of the cosmic microwave background, and suffers from much less ambiguities
concerning unknown physics and/or processes compared to astrophysical and inflationary
mechanisms for seed-field generation. Thus our results are robust.

\section{Formulation}

Euler equations for proton fluid and electron fluid are given by
\begin{eqnarray}
&& ( \delta^{i}_{~\mu} + u^{i} u_{\mu} )
   \left( T^{\mu\nu}_{p ~ ;\nu} + T^{\mu\nu}_{{\rm EM}p ;\nu} \right)
   = C^{(C)i}_{pe} + C^{(T)i}_{p \gamma},
\label{eq:EOMp} \\
&& ( \delta^{i}_{~\mu} + u^{i} u_{\mu} )
   \left( T^{\mu\nu}_{e ~ ;\nu} + T^{\mu\nu}_{{\rm EM}e ;\nu} \right)
   = C^{(C)i}_{ep} + C^{(T)i}_{e \gamma},
\label{eq:EOMe}
\end{eqnarray}
where $T^{\mu\nu}_{p(e)}$ and $T^{\mu\nu}_{{\rm EM}p(e)}$ are the energy-momentum
tensor of proton (electron) fluid and electromagnetic field coupling to protons (electrons)
current, respectively. Here $\mu, \nu = 0,1,2,3$ and $i = 1,2,3$.
The projection of the divergence of the energy-momentum tensors are computed as
\begin{eqnarray}
&& ( \delta^{i}_{~\mu} + u^{i} u_{\mu} ) T^{\mu\nu}_{~~;\nu} \nonumber \\
&& ~~~~
   = ( \rho + p ) u^{\mu} u^{i}_{~;\mu}
     + ( \delta^{i}_{~\mu} + u^{i} u_{\mu} ) P^{,\mu},
\\
&& ( \delta^{i}_{~\mu} + u^{i} u_{\mu} ) T^{\mu\nu}_{{\rm EM} ;\nu}
   = - j^{\nu} F^{\mu}_{~~\nu},
\end{eqnarray}
where $\rho$, $P$ and $j^{\mu}$ are the energy density, pressure and electric current,
respectively. The r.h.s. of Eq. (\ref{eq:EOMp}) and (\ref{eq:EOMe}) represent the collision
terms. $C^{(C)\mu}_{pe} = - C^{(C)\mu}_{ep}$ is the collision term for the Coulomb scattering
between protons and electrons. This term leads to the diffusion of magnetic field
and can be neglected in the highly conducting medium in early universe.
On the other hand, the collsion terms for the Thomson scattering between protons (electrons)
and photons are expressed as $C^{(T)\mu}_{p(e) \gamma}$. Because the collision term for
the protons can be neglected compared to that for the electrons, difference in velocities
of protons and electrons will be induced which leads to electric current. This electric
current becomes a source for magnetic field.

Now we evaluate the collision term for the Thomson scattering:
\begin{equation}
\gamma(p_{i}) + e^{-}(q_{i}) \rightarrow \gamma(p'_{i}) + e^{-}(q'_{i}),
\end{equation}
where the quantities in the parentheses denote the particle momenta.
The collision integral $C^{(T)}_{\gamma e}[f(p_{i})]$ for this scattering is given as
\begin{widetext}
\begin{eqnarray}
C^{(T)}_{\gamma e}[f(p_{i})]
&=& \frac{2 \pi^{4}}{p}
    \int \frac{d^{3}p'}{(2 \pi)^{3} 2 E_{\gamma}(p')}
    \int \frac{d^{3}q}{(2 \pi)^{3} 2 E_{e}(q)}
    \int \frac{d^{3}q'}{(2 \pi)^{3} 2 E_{e}(q')} \nonumber \\
& & \times |M|^{2}
    \delta \left[ E_{\gamma}(p) + E_{e}(q) - E_{\gamma}(p') - E_{e}(q') \right]
    \delta^{(3)} \left[ p_{i} + q_{i} - p'_{i} - q'_{i} \right]
    \left\{ f_{\gamma}(p'_{i}) f_{e}(q'_{i}) - f_{\gamma}(p_{i}) f_{e}(q_{i}) \right\}
    \nonumber \\
&\sim& \frac{\pi n_{e}}{4 m_{e}^{2} p}
    \int \frac{d^{3}p'}{(2 \pi)^{3} p'} |M|^{2}
    \left\{ f_{\gamma}(p'_{i}) - f_{\gamma}(p_{i}) \right\}
    \left\{ \delta (p-p')
            + ( p_{i} - p'_{i} ) u_{e}^{i} \frac{\partial \delta (p-p')}{\partial p'}
    \right\},
\end{eqnarray}
\end{widetext}
where $E_{\gamma(e)}$ and $f_{\gamma(e)}$ are the energy and the distribution function of
photons (electrons), respectively, $|M|^{2}$ is the scattering amplitude and $m_{e}$ is
the electron mass. Then we obtain the collision term in the Euler equation (\ref{eq:EOMe}), as
\begin{eqnarray}
C^{(T)i}_{\gamma e}[f(p_{i})]
&=& \int \frac{d^{3}p}{(2 \pi)^{3}} p^{i} C^{(T)}_{\gamma e}[f(p_{i})] \nonumber \\
&=& \frac{4 \sigma_{T} \rho_{\gamma} n_{e}}{3}
    \left[ ( u_{e}^{i} - u_{\gamma}^{i} ) + \frac{1}{8} u_{ej} \Pi_{\gamma}^{ij} \right],
\label{eq:collision_term}
\end{eqnarray}
where $\sigma_{T}$ is the cross section of the Thomson scattering.
Here moments of the distribution functions are given by
\begin{eqnarray}
&& \int \frac{d^{3}p}{(2\pi)^{3}} p f_{\gamma}(p_{i}) = \rho_{\gamma}, \\
&& \int \frac{d^{3}p}{(2\pi)^{3}} p^{i} f_{\gamma}(p_{i})
   = \frac{4}{3} \rho_{\gamma} u_{\gamma}^{i}, \\
&& \int \frac{d^{3}p}{(2\pi)^{3}} p^{i} f_{e}(p_{i})
   = \rho_{e} u_{e}^{i}, \\
&& \int \frac{d^{3}p}{(2\pi)^{3}} p^{-1} p^{i} p^{j} f_{\gamma}(p_{i})
   = \frac{1}{6} \rho_{\gamma} \Pi_{\gamma}^{ij} + \frac{1}{3} \rho_{\gamma} \delta^{ij},
\end{eqnarray}
where $\Pi_{\gamma}^{ij}$ is photon anisotropic stress. It should be noted that
the collision term (\ref{eq:collision_term}) was obtained non-perturbatively
with respect to the cosmological perturbation.


Altogether, the Euler equations for protons and electrons are written as
\begin{eqnarray}
&& m_{p} n u_{p}^{\mu} u_{p;\mu}^{i} - e n u_{p}^{\mu} F_{\mu}^{~i}
   = 0,
\label{eq:EOM_p2} \\
&& m_{e} n u_{e}^{\mu} u_{e;\mu}^{i} + e n u_{e}^{\mu} F_{\mu}^{~i} \nonumber \\
&& ~~
   = - \frac{4 \sigma_{T} \rho_{\gamma} n}{3}
       \left[ ( u_{e}^{i} - u_{\gamma}^{i} ) + \frac{1}{8} u_{ej} \Pi_{\gamma}^{ij} \right],
\label{eq:EOM_e2}
\end{eqnarray}
where $m_{p}$ is the proton mass, and the pressure of proton and electron fluids are neglected.
We also assumed the charge neutrality: $n = n_{e} \sim n_{p}$.
Subtracting Eq. (\ref{eq:EOM_p2})
multiplied by $m_{e}$ from Eq. (\ref{eq:EOM_e2}) multiplied by $m_{p}$, we obtain
\begin{eqnarray}
&& - \frac{m_{p}m_{e}}{e}
   \left[ n u^{\mu} \left( \frac{j^{i}}{n} \right)_{;\mu}
          + j^{\mu}
            \left( \frac{m_{p} - m_{e}}{m_{p} + m_{e}} \frac{j^{i}}{en} - u^{i} \right)_{;\mu}
   \right]
\nonumber \\
&& + e n ( m_{p} + m_{e} ) u^{\mu} F^{i}_{~\mu}
   - ( m_{p} - m_{e} ) j^{\mu} F^{i}_{~\mu} \nonumber \\
&& = - \frac{4 m_{p} \sigma_{T} \rho_{\gamma} n}{3}
       \left[ ( u_{e}^{i} - u_{\gamma}^{i} ) + \frac{1}{8} u_{ej} \Pi_{\gamma}^{ij} \right],
\label{eq:Ohm1}
\end{eqnarray}
where $u^{\mu}$ and $j^{\mu}$ are the center-of-mass 4-velocity of the proton and
electron fluids and the net electric current, respectively, defined as
\begin{eqnarray}
&& u^{\mu} \equiv \frac{m_{p}u_{p}^{\mu} + m_{e}u_{e}^{\mu}}{m_{p} + m_{e}}, \\
&& j^{\mu} \equiv e n (u_{p}^{\mu} - u_{e}^{\mu}).
\end{eqnarray}
Noting the Maxwell equations $F^{\mu\nu}_{;\nu} = j^{\mu}$, we see that
the first term in the l.h.s. of Eq. (\ref{eq:Ohm1}) are suppressed, compared
to the second term, by a factor \cite{Subramanian94}
\begin{equation}
\frac{c^{2}}{L^{2} \omega_{p}^{2}}
\sim 3 \times 10^{-47}
     \left( \frac{10^{10} {\rm cm}^{-3}}{n} \right)
     \left( \frac{1 {\rm Mpc}}{L} \right)^{2},
\end{equation}
where $c$ is the speed of light, $L$ is a characteristic length of the system
and $\omega_{p} = \sqrt{4 \pi n e^{2}/m_{e}}$ is the plasma frequency.
Second order vector perturbations are contained in the covariant derivative of
the electric current and were evaluated in \cite{Matarrese04}. They obtained
the current magnetic field of order $10^{-29} {\rm G}$ at 1Mpc scale.
As we will see below, this is much smaller than a value we obtain in this paper,
which justifies the neglection of the first term in the l.h.s. of Eq. (\ref{eq:Ohm1}).

The third term in the l.h.s. of Eq. (\ref{eq:Ohm1}) is the Hall term which can also
be neglected because the Coulomb coupling between protons and electrons is so tight
that $|u^{i}| \gg |u_{p}^{i} - u_{e}^{i}|$. Then we have a generalized Ohm's law:
\begin{equation}
u^{\mu} F^{i}_{~\mu}
= - \frac{4 \sigma_{T} \rho_{\gamma}}{3 e}
      \left[ ( u_{e}^{i} - u_{\gamma}^{i} ) + \frac{1}{8} u_{ej} \Pi_{\gamma}^{ij} \right]
\equiv C^{i}.
\end{equation}

Now we derive the evolution equation for magnetic field. This can be obtained
from the Bianchi identities $F_{[\mu\nu,\lambda]} = 0$, as
\begin{eqnarray}
0
&=& \epsilon^{ijk} u^{\mu} F_{[jk,\mu]} \nonumber \\
&=& u^{\mu} B^{i}_{~\mu} - \frac{1}{u^{0}} \epsilon^{ijk} C_{j} u^{0}_{~,k}
    + \epsilon^{ijk} C_{j,k} \nonumber \\
& & - ( u^{i}_{~,j} B^{j} - u^{j}_{~,j} B^{i} )
    + \frac{u^{0}_{~,j}}{u^{0}} ( B^{j} u^{i} - B^{i} u^{j} ),
\label{eq:Bianchi}
\end{eqnarray}
where $\epsilon^{ijk}$ is the Levi-Civit\`{a} tensor and
$B^{i} \equiv \epsilon^{ijk} F_{jk}/2$ is comoving magnetic field. We can expand
the photon energy density, fluid velosities and photon anisotropic stress as
\begin{eqnarray}
&& \rho_{\gamma} = \st{0}{\rho}_{\gamma} + \st{1}{\rho}_{\gamma} + \cdots, ~~~
   u_{0} = 1 + \st{2}{u}_{0} + \cdots, \nonumber \\
&& u_{i} = \st{1}{u}_{i} + \st{2}{u}_{i} + \cdots, ~~~
   \Pi_{\gamma}^{ij} = \st{1}{\Pi}_{\gamma}^{ij} + \cdots,
\end{eqnarray}
where the superscripts $(i)$ denote the order of expansion. Remembering that $B^{i}$
is a small quantity, we see that most of the terms in Eq. (\ref{eq:Bianchi}),
other than the first and third terms, can be neglected. Thus we obtain
\begin{eqnarray}
\dot{B}^{i}
&\sim& - \epsilon^{ijk} C_{j,k} \nonumber \\
&\sim& \frac{4 \sigma_{T} \st{0}{\rho}_{\gamma}}{3 e} \epsilon^{ijk}
       \Biggl[ \frac{\st{1}{\rho}_{\gamma ,k}}{\st{0}{\rho}_{\gamma}} 
              \left( \st{1}{u}_{ej} - \st{1}{u}_{\gamma j} \right)
              + \left( \st{2}{u}_{ej,k} - \st{2}{u}_{\gamma j,k} \right) \nonumber \\
&&              + \frac{1}{8} \left( \st{1}{u}_{el,k} \st{1}{\Pi}{}^{l}_{\gamma j}
                                   + \st{1}{u}_{el} \st{1}{\Pi}{}^{l}_{\gamma j,k} \right)
       \Biggr],
\label{eq:B_dot}
\end{eqnarray}
where the dot denotes a derivative with respect to the cosmic time, and we used the fact
that there is no vorticity in the linear order: $\epsilon^{ijk} \st{1}{u}_{j,k} = 0$.
The contributions of the first two terms in Eq. (\ref{eq:B_dot}) were first noticed
in \cite{Gopal04}. From this expression, we see that magnetic field cannot be generated
in the linear order. Here it should be noted that the velocity of electron fluid
can be approximated to the center-of-mass velocity at this order,
$\st{1}{u}_{e}^{i} \sim \st{1}{u}^{i}$.

The linear-order quantities have only scalar components so that we can write as
\begin{eqnarray}
&& \st{1}{u}_{e}^{i} \sim \st{1}{u}^{i} = - i \hat{k}^{i}  u, ~~~~
   \st{1}{u}_{\gamma}^{i} = - i \hat{k}^{i} u_{\gamma}, \\
&& \st{1}{\Pi}{}^{i}_{\gamma j}
   = - ( \hat{k}^{i} \hat{k}_{j} - \frac{1}{3} \delta^{i}_{~j} ) \Pi_{\gamma},
\end{eqnarray}
 Also we define
$\bar{\rho}_{\gamma} \equiv \st{0}{\rho}_{\gamma}$ and
$\delta_{\gamma} \equiv \st{1}{\rho}_{\gamma} / \bar{\rho}_{\gamma}$.
Then we can rewrite the Fourier component of $B^{i}$ in terms of the Fourier components
of $\delta_{\gamma}$, $u$, $u_{\gamma}$ and $\Pi_{\gamma}$, as
\begin{widetext}
\begin{eqnarray}
B^{i}(K,t)
&=& \frac{4 \sigma_{T}}{3 e} \epsilon^{ijk} \int^{t} dt' \bar{\rho}_{\gamma}(t')
    \int \frac{d^{3}k'}{(2\pi)^{3}} k'_{j} K_{k}
         \bigg[ \delta_{\gamma}(|\vec{K}-\vec{k}'|,t') \delta u_{b\gamma} (k',t')
                - \frac{3 \vec{K} \cdot \vec{k}' - 2 k'{}^{2}}{24 k'{}^{2} |\vec{K}-\vec{k}'|}
                  u (|\vec{K}-\vec{k}'|,t') \Pi_{\gamma} (k',t')
         \bigg] \nonumber \\
& & - \frac{4 \sigma_{T}}{3 e}
      \int^{t} dt' \bar{\rho}_{\gamma} (t')
      \left[ \Omega^{i} (K,t') - \Omega_{\gamma}^{i} (K,t') \right],
\label{eq:B_Fourier}
\end{eqnarray}
\end{widetext}
where $\hat{k}_{i}$ is the unit vector in direction of $k_{i}$,
$\delta u \equiv u - u_{\gamma}$. Vorticities, which come from the second-order
vector perturbation, are defined as $\Omega^{i} = \epsilon^{ijk} \st{2}{u}_{k,j}$ and
$\Omega_{\gamma}^{i} = \epsilon^{ijk} \st{2}{u}_{\gamma k,j}$.
Eq. (\ref{eq:B_Fourier}), which describes the evolution of magnetic field, is one of
our main results.

\section{Evaluation of source terms}

\begin{figure}
\centering
  \rotatebox{0}{\includegraphics[width=0.42\textwidth]{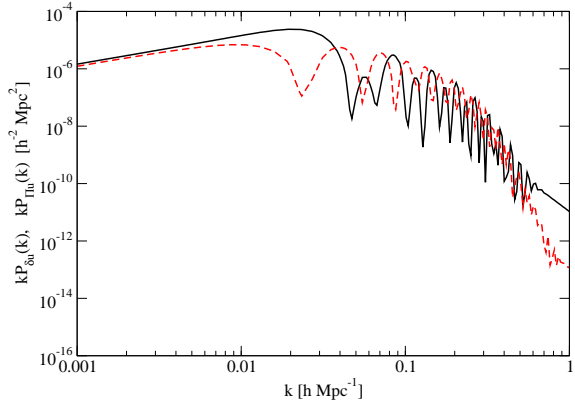}}
  \caption{Power spectra (multiplied by $k$)  in the source term of
 Eq.(\ref{eq:B_Fourier}) at $z \approx 1100$. The solid and dashed lines
 correspond to power spectra of
 $\left<|\delta_{\gamma} (u - u_{\gamma})| \right>$ and
 $\left<|\Pi_{\gamma} u| \right>$, respectively. The spectra
 have been normalized to agree on large scales at present epoch.}
\label{fig1}
\end{figure}

Here we briefly evaluate the first two terms in the r.h.s. of Eq. (\ref{eq:B_Fourier})
which are constructed from the first-order quantities. The evaluation can be easily
accomplished by using linearized Boltzmann code such as CMBFAST \cite{cmbfast}.
In figure \ref{fig1} we show the power spectra of these terms multiplied by $k$,
$k P_{\delta u}(k)$ and $k P_{\Pi u}(k)$, around the last scattering epoch
$z \approx 1100$. Here the power spectra are defined as
\begin{eqnarray}
&& \left< |\delta_{\gamma}(\vec{k}) \{u - u_{\gamma}\}(\vec{k}')| \right>
   = P_{\delta u}(k) \delta^{(3)}(\vec{k} - \vec{k}') \\
&& \left< |\Pi_{\gamma}(\vec{k}) u(\vec{k}')| \right>
   = P_{\Pi u}(k) \delta^{(3)}(\vec{k} - \vec{k}').
\end{eqnarray}
For this illustration all cosmological parameters were
fixed to the $\Lambda$CDM values, i.e., $(h,n_s,\Omega_b
h^2,\Omega_\Lambda,A)=(0.71,0.99,0.022,0.74,0.83)$.

We found that these two terms contribute to the generation of magnetic
field at almost the same order of magnitude at all scales. This can be understood since
$\delta_{\gamma} \sim u_{\gamma}$ (acoustic oscillations),
$u - u_{\gamma} \sim \varepsilon R u$, and
$\Pi_{\gamma} \sim \varepsilon u_{\gamma}$ in the tight coupling approximation
of baryon-photon fluid, where $\varepsilon$ being tight coupling parameter
and $R = 3 \rho_{b} / 4 \rho_{\gamma} \sim {\cal O}(1)$ around the last scattering epoch.
From the spectra we can estimate the field strength of generated magnetic field
at the decoupling epoch as
$B \sim \frac{8 \sigma_{T} a \rho_{\gamma}}{3e} k^{3} P_{\delta u}(k)
   \sim 10^{-19} {\rm G}$ at 10Mpc comoving scale. If the magnetic field
decays adiabatically according to the cosmic expansion, the current magnetic
field is about $10^{-25} {\rm G}$.

\section{Discussion and summary}

In this letter, we discussed generation of magnetic field from cosmological perturbations.
Separate treatment of protons, electrons and photons, and appropriate evaluation
of collision term allowed us to obtain the evolution equation for magnetic field.
We saw that electric current would be induced by collision term between electrons
and photons. Up to the second order, there are three contributions which act as sources
for magnetic field. In addition to well-known vorticity effect, we found a contribution
from the anisotropic stress of photons. Then we estimated the magnitudes of two of
the three contributions which can be evaluated from the first-order quantities.
We showed the two contributions are comparable at all order and the amplitude of
the produced magnetic field is about $10^{-19} {\rm G}$ at 10Mpc comoving scale
at the decoupling.
Concerning these two contributions, power spectrum of the produced magnetic fields
and correlation with the cosmic microwave background must be studied.
We will present them elsewhere soon \cite{Ichiki05}.

To evaluate the vorticity term, second-order perturbation theory is necessary. Although
many authors have been trying this
\cite{Bruni97,Matarrese98,Nakamura:2004wr,Tomita05},
the complete formulation including matter perturbations is still a challenging problem.

K. T. and K. I. are supported by Grant-in-Aid for JSPS Fellows.

\end{document}